\documentclass[aps,prl,twocolumn,superscriptaddress]{revtex4}
\usepackage{amsmath,epsfig,color,amssymb}
\usepackage{graphicx}
\usepackage{graphicx}
\usepackage{dcolumn}
\usepackage{bm}
\usepackage{graphicx}
\usepackage{amsmath, amsfonts, amssymb,mathrsfs}
\usepackage{pstricks}
\usepackage{amsxtra}
\usepackage{amsthm}
\usepackage{natbib}

\def\be{\begin{equation}}
\def\ee{\end{equation}}
\def\bea{\begin{eqnarray}}
\def\eea{\end{eqnarray}}
\newcommand{\ket}[1]{\mbox{$|#1\rangle$}}
\newcommand{\bra}[1]{\mbox{$\langle#1|$}}

\def\be{\begin{equation}}      
\def\ee{\end{equation}}
\def\beu{\begin{equation*}}   
\def\eeu{\end{equation*}}

\providecommand{\abs}[1]{\left\lvert#1\right\rvert}   
\providecommand{\ket}[1]{\left|#1\right\rangle}

\providecommand{\bra}[1]{\left\langle#1\right|}
\providecommand{\mean}[1]{\left\langle#1\right\rangle}

\definecolor{new}{rgb}{.08,.05,.8}

\newcommand{\delete}[1]{{}}

\graphicspath{}

\begin{document}
\title{Sisyphus Thermalization of Photons in a Cavity-Coupled Double Quantum Dot}
\author{M.~J.~Gullans}
\affiliation{Joint Quantum Institute, National Institute of Standards and Technology, Gaithersburg, Maryland 20899, USA}
\affiliation{Joint Center for Quantum Information and Computer Science, University of Maryland, College Park, Maryland 20742, USA}
\author{J. Stehlik}
\author{Y.-Y. Liu}
\author{C. Eichler}
\affiliation{Department of Physics, Princeton University, Princeton, New Jersey 08544, USA}
\author{J. R. Petta}
\affiliation{Department of Physics, Princeton University, Princeton, New Jersey 08544, USA}
\author{J.~M.~Taylor}
\affiliation{Joint Quantum Institute, National Institute of Standards and Technology, Gaithersburg, Maryland 20899, USA}
\affiliation{Joint Center for Quantum Information and Computer Science, University of Maryland, College Park, Maryland 20742, USA}

\begin{abstract}
We investigate the non-classical states of light that emerge in a microwave resonator coupled to a periodically-driven electron in a nanowire double quantum dot (DQD).
Under certain drive configurations, we find that the resonator approaches a thermal state at the temperature of the surrounding substrate with a chemical potential given by a harmonic of the drive frequency.  Away from these thermal regions we find regions of gain and loss, where the system can lase, or regions where the DQD acts as a single-photon source.
These effects are observable in current devices and have broad utility for quantum optics with microwave photons.
%
%
\end{abstract}
\maketitle


When a physical system is subject to periodic driving, the usual notions of equilibrium thermodynamics have to be revisited.  For a closed system, the second law of thermodynamics suggests it approaches an infinite temperature state; however, there are dramatic exceptions to this behavior in integrable systems \cite{Kapitza51,Lieberman72,Casati79,Grempel84,Graham92,Moore94,Lemarie09} and the recently discovered class of many-body localized phases \cite{Basko06,Pal10,DAlessio13,Lazarides15,Ponte15}.  For open systems, where the periodically driven system is  coupled to a fixed temperature bath, the system naturally reaches a steady state that evolves with the same periodicity as the drive; however, unlike in thermal equilibrium, no general classification scheme is believed to exist for such states \cite{Kohn01,Kitagawa11,Lindner11,Wang13,Langemeyer14,Shirai15,Liu15}.  


Solid-state qubits are a versatile platform to study strongly driven quantum systems \cite{Oliver05,Sillanpaa06,Nori10,Petta10,Gaudreau12,Stehlik12,Deng15,Childress10,Fuchs11}.
In the case of gate-defined quantum dots and superconducting qubits, the typical energy splittings are  small enough that  the drive amplitude can be much larger than the qubit splitting \cite{Nori10,Petta10,Gaudreau12,Stehlik12}.  
When  superconducting qubits are integrated in a circuit quantum electrodynamics (cQED) architecture and strongly driven, they can be used to generate non-classical states of light \cite{Houck07,Wallraff10}, lasing \cite{Grajcar08}, and thermal states of light with a chemical potential \cite{Hafezi14}. For cQED with quantum dots, theoretical and experimental work has focused on weak driving or incoherent tunneling through the leads \cite{Childress04,Petersson12,Frey12,Liu14,Liu15,Stockklauser15}; however, the effect of strong driving remains unexplored.  For similar drive parameters, we expect qualitatively different behavior from superconducting qubits because of the strong electron-phonon coupling in quantum dots \cite{Fujisawa98,Petta04,Gullans15}.


In this Letter, we investigate a microwave resonator interacting with a periodically driven electron in a double quantum dot (DQD).  The DQD is coupled to a fixed temperature phonon bath. We investigate the non-classical states of light that emerge in the long time limit.  For certain drive configurations, the resonator field approaches a thermal state at the phonon temperature with a chemical potential given by a harmonic of the drive frequency.  Away from these thermal regions, we find regions where  the system begins lasing or acts as a single-photon source.


%
%
%


We take the DQD to be configured near the charge transition between the states $\ket{L}$ and $\ket{R}$ with one electron in either the left or right dot, respectively.  This pair of states has a large electric dipole moment that couples to the electric field in a nearby microwave resonator, as well as acoustic phonons in the semiconductor host [see Fig.~1(a)] \cite{Childress04,Petersson12,Frey12}. In this work, we focus on the case of an InAs nanowire DQD, as realized in recent experiments \cite{Petersson12}; however, many of the results apply to other DQD-cQED systems under study \cite{Frey12,Stockklauser15}.   

\begin{figure}[b]
\begin{center}
\includegraphics[width = .45 \textwidth]{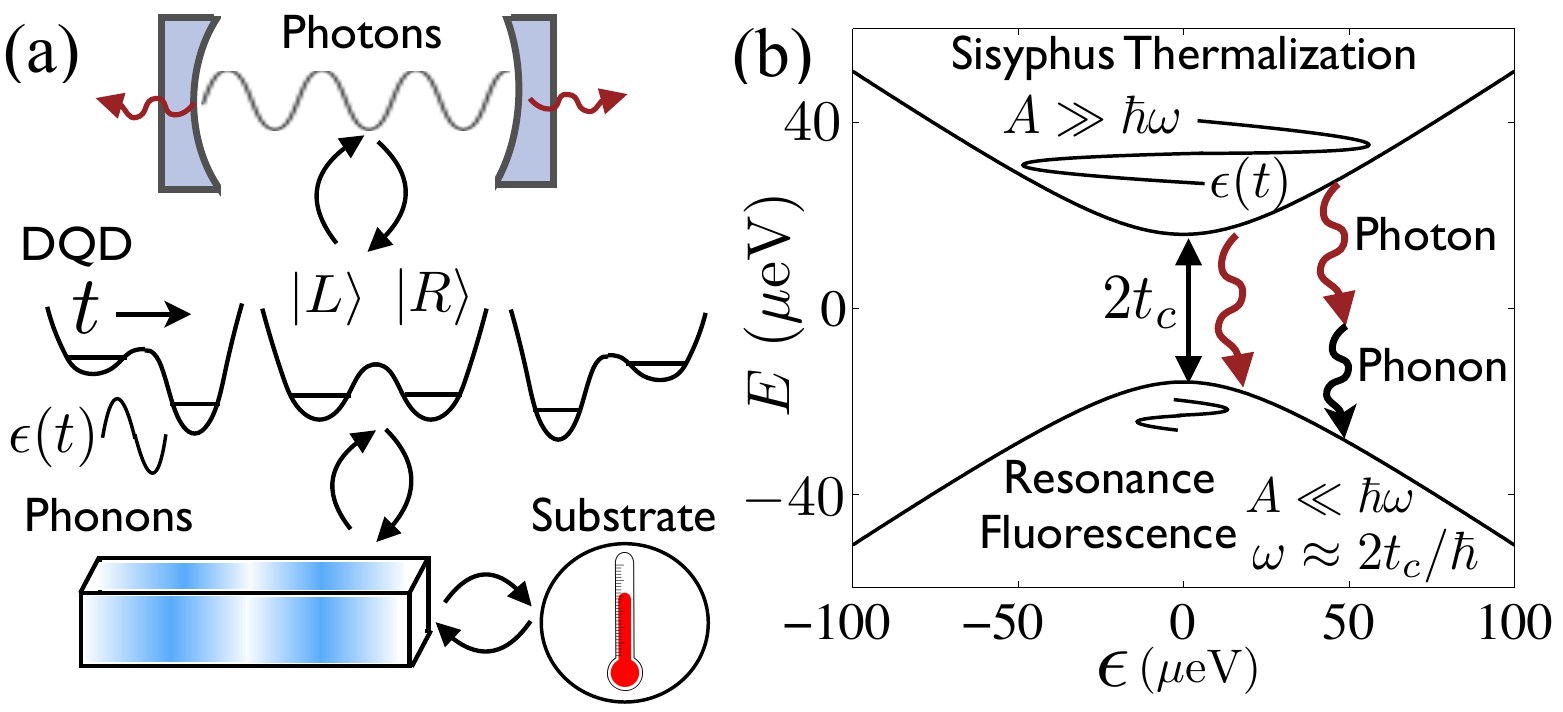}
\caption{(a) A microwave resonator is coupled to charge states in a DQD. The DQD is subject to periodic driving and strongly coupled to acoustic phonons, which are held at a fixed temperature. (b) Energy level diagram of the DQD with $t_c = 20~\mu$eV.  For  $A \ll \hbar \omega$ with  $\omega \approx 2 t_c/\hbar$, resonance fluorescence of the DQD dominates, leading to antibunched light.  When $A\gg \hbar \omega$,  the phonon sideband dominates, leading to thermalization.} 
\label{default}
\end{center}
\end{figure}

We consider periodic driving of the level detuning $\epsilon(t)=\epsilon_0+   A \cos \omega t$, where $\epsilon_0$ is an offset detuning and $A$ and $\omega$ are the amplitude and frequency of the drive.  In a process reminiscent of Sisyphus from Greek mythology \cite{Grajcar08}, the DQD is continually excited by the drive, only to relax to the ground state via phonon and photon emission [see Fig.~1(b)].
For low driving amplitudes, $A \ll \omega$ with a drive that is near resonant with the DQD,  the photon dynamics are dominated by resonance fluorescence, where the DQD acts as a single-photon source \cite{Kimble76,Kimble77}. 

When $A \gtrsim \hbar \omega$, the situation changes dramatically because the two-level nature of the DQD leads to a series of harmonics  (denoted by index $n$) of the drive frequency  up to $ n_\textrm{max}  \approx A/\hbar \omega$ \cite{Grifoni98,Oosterkamp98}.  
These sidebands give rise to a parametric ``time-varying'' coupling between resonator photons and phonons mediated by the DQD.  
In the absence of other processes, such a parametric coupling of photons to a thermal bath can lead to thermal light with a chemical potential by equilibrating the photons with low frequency bath modes \cite{Hafezi14}.


 Solving for the long-time dynamics using Floquet analysis, we uncover several regimes where the resonator photons approach a thermal state in the strongly-driven limit. 
In the Floquet picture, where energy is only well-defined modulo $\hbar \omega$ \cite{Sambe73},  the resonator frequency $\omega_c$ is mapped onto $\delta= \omega_c - n_c \omega$, where  $n_c$ is the  closest integer to $\omega_c/\omega$.
Crucially, near the resonances $\delta = 0$, the photon dynamics become dominated by Raman scattering events in which the DQD absorbs $n_c$ drive quanta while simultaneously annihilating or creating a resonator photon and a phonon at frequency $\abs{\delta}$.   In a process we refer to as ``Sisyphus thermalization,'' these scattering events thermalize the resonator to the substrate temperature witha a chemical potential $\mu= \hbar n_c \omega$.  This effect is enhanced in InAs nanowires because the phonon spectral density for piezoelectric coupling to the DQD $\mathcal{J}(\nu) \sim \nu$ for small $\nu$, as compared to, e.g., GaAs DQDs where $\mathcal{J}(\nu) \sim \nu^3$ \cite{Brandes05,Weber10}.  Away from these thermal regions, we find regions of gain and loss, where the system can begin lasing \cite{Grajcar08}, as well as
 regimes more consistent with resonance fluorescence, where the DQD acts as a single-photon source \cite{Kimble76,Kimble77}.

\emph{Floquet Model.--}
The Hamiltonian for the periodically driven DQD takes the form
\begin{align}\label{eqn:H}
H_c(t)&= \frac{1}{2} \big( \epsilon_0 + A \cos \omega t \big) \sigma_z + t_c \sigma_x 
\end{align}
where $\sigma_\nu$ are Pauli matrices operating in the $\{\ket{L},\ket{R}\}$ orbital subspace, and $t_c$ is the tunnel coupling between the dots.
Including the resonator and phonons
\begin{align} \label{eqn:H}
H&=H_c(t) +  \hbar \omega_c a^\dagger a + \sum_k \hbar \omega_k a_k^\dagger a_k + \hat{X} \sigma_z, 
\end{align}
where   $\omega_k$ is the frequency of the $k$th phonon mode, $a(a^\dagger)$ and $a_k(a_k^\dagger)$ are the bosonic photon and phonon annihilation (creation) operators, respectively, $\hat{X}/\hbar=g_c (a+ a^\dagger) + \sum_k \lambda_k (a_k^\dagger+a_k)$ contains the coupling $g_c$ between the resonator and DQD, as well as the coupling $\lambda_k$ between the DQD and the phonons.

From Floquet theory \cite{Sambe73,Grifoni98}, we know that the evolution under such time-periodic Hamiltonians can be  formally represented using an infinite dimensional basis $\ket{m}$ for integers $m$.  In this representation, $H(t)$ is mapped to a time-independent Hamiltonian $H_F$ by 
adding the term $\hbar \,\omega \hat{N} =\sum_m \hbar\, \omega\, m \ket{m}\bra{m}$ and 
converting the functions $e^{-i m \omega t}$  into  operators  which change the Floquet index by $m$, i.e., $\hat{F}_m = \sum_{m'} \ket{m+m'}\bra{m'}$.

Before writing $H_F$, we  apply three unitary transformations that map the problem to a convenient basis.
First, we apply a polaron transformation that dresses the DQD with the ambient phonons and photons in the environment  and gives to explicit interaction terms between photons and phonons \cite{Brandes05,Gullans15}
\be
U_p = e^{ \big[ g_c  (a- a^\dagger)/\omega_c + \sum_{k} \lambda_k (a_{k}-a_{k}^\dagger)/\tilde{\omega}_k \big] \sigma_z },
\ee
where we have defined the renormalized $\tilde{\omega}_k = \sqrt{\omega_k^2 + \eta^2}$ to regularize the infrared divergences  near $\omega_k =0$. This regularization is consistent with our assumption that the phonon bath is coupled to a fixed temperature reservoir, where $\eta$ is the thermalization rate of the phonons.   
In this treatment, the thermodynamic limit corresponds to taking $\eta \to 0$, while maintaining the nanowire phonons at a fixed temperature.
The second transformation, motivated by the strong periodic driving, folds the resonator and phonon spectrum into a band between $\pm \hbar \omega/2$
\be
U_{RW}(t) = e^{i n_c \omega a^{\dagger} a t} \prod_{n\ge0 }\prod_{k\in \Omega_n} e^{  i n \omega a_k^\dagger a_k t},
\ee
where 
$\Omega_n$ is the set of $k$ such that $(n-1/2)\omega<\omega_k<(n+1/2)\omega$.  
Finally, 
we apply a unitary $U_F$ (found numerically) such that $U_F H_c^F U_F^\dagger$ is diagonal, where $H_c^F$ is the Floquet Hamiltonian associated with $H_c$.  This modifies the coupling between the DQD and the photons and phonons through the tensor $u_{\nu m}^{\mu m}$ defined by 
\be
U_F \sigma_\nu \hat{F}_n U_F^\dagger = \sum_{\mu\in \{x,y,z\}}\sum_{m=-\infty}^{\infty} u_{\nu n}^{\mu m} \sigma_\mu \hat{F}_m.
\ee
With these transformations the Floquet Hamiltonian, to lowest order in $g_c/\omega_c$ and $\lambda_k/\omega_k$, is
\begin{align} \label{eqn:hf}
\frac{H_F}{\hbar}&=\frac{\Delta}{2} \sigma_z + \delta a^\dagger a +\sum_{n,k\in \Omega_n}  (\omega_k - n \omega) a_k^\dagger a_k + \omega \hat{N},  \\ \nonumber
&+ (\hat{V}_{cp}+\hat{V}_{cpp} + h.c.),  \\  \label{eqn:vcp}
\hat{V}_{cp}&= \sum_{n,\mu,m} u_{y n}^{\mu m}  \bigg(  \frac{2 i  t_c g_c}{\hbar \omega_c}   \delta_{nn_c} a + \hat{P}_n \bigg) \sigma_\mu \hat{F}_{m} \\  \label{eqn:vcpp}
\hat{V}_{cpp}&= \sum_{n,\mu,m} \frac{g_c \hat{P}_n}{\omega_c} (u_{x n+nc}^{\mu m} a + u_{x n-nc}^{\mu m} a^\dagger) \sigma_\mu \hat{F}_m,
\end{align}
where $\pm\hbar \Delta/2$ are the two quasi-energies of $H_c^F$, $\delta_{nn_c}$ is the Kronecker delta function, $\hat{P}_n = \sum_{k\in \Omega_n} \frac{2 i t_c \lambda_k}{\hbar \tilde{\omega}_k} a_k$, and the summation limits are the same as above.   The  term $V_{cp}$ describes first order DQD-resonator and DQD-phonon interactions, while $V_{cpp}$ describes simultaneous phonon-resonator-DQD interactions.   When $A \gg t_c^2/\hbar \omega$,   $ \hbar \Delta \approx \sqrt{\epsilon_0^2+4 t_c^2 J_0^2(A/\hbar \omega)}$, where $J_n$ refers to the Bessel functions arising from the identity $e^{i A \sin \omega t/2 \hbar \omega}= \sum_n J_n(A/\hbar \omega) e^{i n \omega t}$ \cite{Nori10}.  


\emph{Steady State.--}
We have chosen a basis where the first four terms in $H_F$, given in Eq.~(\ref{eqn:hf}), are diagonal; however, $V_{cp}$ and $V_{cpp}$ are not diagonal and  will lead to a slow time evolution in this basis.  We now solve for this effective time-evolution in the Floquet basis using standard methods from the theory of open quantum systems.
We use a basis ordering convention such that $\Delta < \omega/2$. 
In this basis, single-phonon processes can only resonantly couple states with the same Floquet index. As a result, the Floquet blocks evolve approximately independently from each other.
The rate to spontaneously emit a phonon in the $\Omega_n$ band 
and make a transition from the upper (lower) to the lower (upper)  state is 
\be
\gamma_{n\mp} = \frac{8 \pi t_c^2/\hbar^2}{(n \omega \pm \Delta)^2}(\lvert{u_{y n}^{x 0}}\lvert^2 
+\lvert{u_{y n}^{y 0}}\lvert^2)\, \mathcal{J}( n \omega \pm \Delta),
\ee
where  $\mathcal{J}(\nu)= \sum_k \abs{\lambda_k}^2 \delta(\nu - \omega_k)$ is the phonon spectral density.  We assume the phonons are in thermal equilibrium with temperature $T$ and distributed according to the Bose function $n_p(\nu)=(e^{ \hbar \nu/k_B T}-1)^{-1}$.  In this case, there is also stimulated emission and absorption at the rates $\gamma_{n\mp}^s = \gamma_{n\mp} \, n_p(n\omega \pm \Delta) $.  
 We find the DQD states within each Floquet block follow the master equation
\be \label{eqn:ME}
\dot{\rho}_n = i \frac{\Delta}{2} [ \sigma_z ,\rho_n] + \gamma_- \mathcal{D}[\sigma_-]\rho_n + \gamma_+ \mathcal{D}[\sigma_+]\rho_n, 
\ee
where $\gamma_{\mp} = \sum_n \gamma_{n\mp}+\gamma_{n\mp}^s+\gamma_{n\mp}^s$ is the total transition rate from the upper (lower) to the lower (upper) DQD states,  $\mathcal{D}[c]\rho_n= -1/2 \{c^\dagger c,\rho_n\} + c \rho_n c^\dagger$ and the total density matrix for the DQD is $\rho_{d} = \sum_n \rho_n \ket{n}\bra{n}$.   This master equation can be used to derive the steady state and all  time-dependent correlation functions of the DQD \cite{supp}.  When $ t_c g_c/\hbar \omega_c \ll \Delta$,  we can use these correlation functions to determine the evolution of the resonator because it has negligible back-action on the DQD.


\emph{Sisyphus Thermalization.--}
Based on the discussions above, for finite $g_c$, we expect three possible types of output light.  When resonance fluorescence dominates, the DQD acts as a single photon source and produces anti-bunched light.  When $\mean{\sigma_\nu} \ne 0$ for some $\nu$, or when the system begins lasing \cite{Liu15},  the DQD will drive the resonator into a coherent state. Finally, if the DQD mostly acts to thermalize the resonator, the light will exhibit thermal statistics.

Conveniently, the four-point correlation function 
\be \label{eqn:g2}
g^{(2)}(\tau) = \lim_{t \to \infty}\frac{\mean{a^\dagger(t)a^\dagger(t+\tau)a(t+\tau) a(t)}}{\mean{a^\dagger(t) a(t)}},
\ee
can distinguish these three states because $g^{(2)}(0)$ equals zero for anti-bunched light, one for a coherent state, and two for thermal light.  
Figure 2(a) shows $g^{(2)}(0)$ for a nanowire DQD, calculated following the approach detailed in the supplementary material \cite{supp}, over a large range of $A$ and $\omega$.  
The parameters defining $\mathcal{J}(\nu)$ are based on recent experiments in InAs nanowires \cite{Liu14}.  We took a separation between the two dots of 120~nm, a longitudinal  confinement of 25~nm for each dot,  a phonon speed of sound of $4\, 000$~m/s,  and the DQD relaxation rate at zero detuning to be  6~ns$^{-1}$ \cite{Brandes05,Petta04, Weber10}.  

\begin{figure}[t]
\begin{center}
\includegraphics[width = .4 \textwidth]{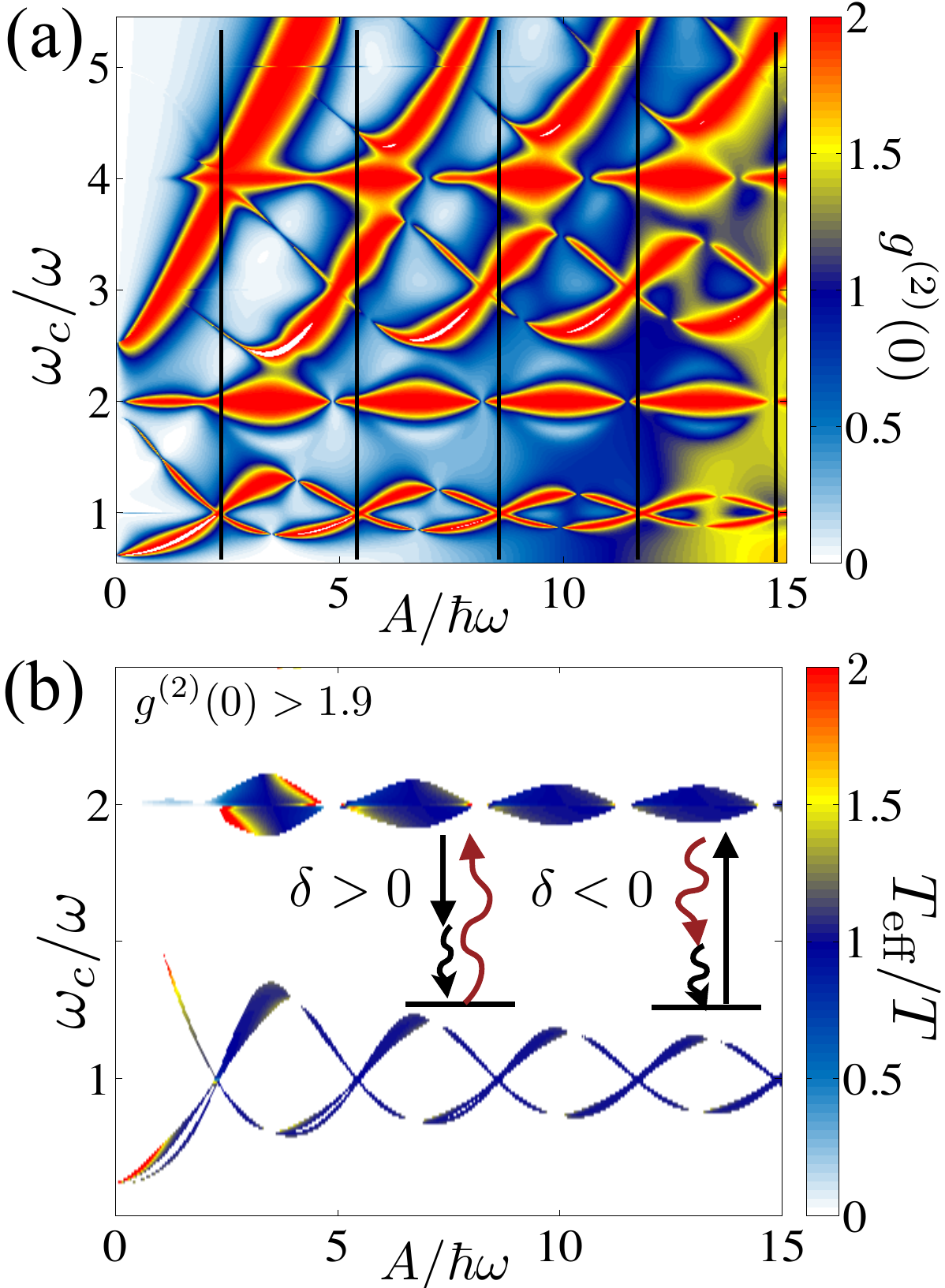}
\caption{(a)  $g^{(2)}(0)$ as a function of $A$ and $\omega$.  We took $\omega_c/2\pi=16$~GHz, $t_c=20~\mu$eV, $\epsilon_0=0$, $g_c/2\pi =70~$MHz, $\kappa/2\pi = 1.3$ MHz, $\eta=0.5$~ns$^{-1}$, and $T=200$~mK.  
Vertical lines correspond to zeros of $J_0(A/\hbar \omega)$. (b) Effective temperature of the photons in the thermal regions (defined by $g^{(2)}(0)>1.9$), with cavity decay neglected.  (Inset) Raman scattering processes leading to thermalization of photons when $\delta=\omega_c -n_c \omega$ is near zero.  The red line is a photon, the solid line is the drive, and the curved black line is a phonon.  }
\label{default}
\end{center}
\end{figure}

Although the behavior of $g^{(2)}(0)$ shown in Fig.~2(a) is a complex function of the drive parameters, we can identify several general features.  First, the singular points in  $g^{(2)}(0)$ are correlated with the zeros of  the quasi-energy $\hbar\Delta$, which, for $\epsilon_0=0$, occurs roughly at the zeros of  $J_0(A/\hbar \omega)$.
 Second, there are large regions where the light has mostly thermal correlations, which tend to occur when $\omega_c/\omega$ is near an integer $n_c$.  
Away from these thermal resonances, the resonator is either strongly antibunched or in a complex, mixed state of thermal, antibunched, and coherent light.  

To better understand how these thermal regions emerge note that, near these resonances, the photon dynamics are dominated by  incoherent Raman scattering processes in which both a photon and phonon are absorbed or emitted without changing the state of the DQD (see Fig.~2(b) inset).  This occurs because the spectral density for these Raman processes near these resonances,
\be
\mathcal{J}_R(\delta) = \sum_k \frac{4 t_c^2 g_c^2 \abs{\lambda_k}^2}{\hbar^2 \tilde{\omega}_k^2 \omega_c^2 }\delta(\delta-\omega_k) = \frac{ 4 t_c^2 g_c^2}{\hbar^2 \omega_c^2}\, \frac{\mathcal{J}(\delta)}{ \delta^2+ \eta^2},
\ee
 diverges for small $\eta$ for an InAs nanowire DQD (where $\mathcal{J}(\nu) \sim \nu$ \cite{Weber10}).  As a result, the photons follow a simple master equation \cite{supp}
\be \label{eqn:rhoc}
\dot{\rho_c}  = i \delta [a^\dagger a, \rho_c]+ (\kappa+R_a) \mathcal{D}[a] \rho_c + R_e \mathcal{D}[a^\dagger] \rho_c,
\ee
where $\kappa$ is the resonator decay rate and $R_{a(e)}$ are the phonon-assisted, photon annihilation (creation) rates, respectively.  When $R_e \ne 0$ and $R_e < \kappa + R_a$, this master equation always leads to a thermal distribution with $g^{(2)}(0) = 2$ \cite{QuantumOpticsBook}.

To see how the chemical potential emerges, we have to consider the regimes $\omega > \omega_c/n_c$ and $\omega<\omega_c/n_c$ separately.  When $ \omega < \omega_c/n_c$ the dominant processes are ones in which a photon is created (annihilated) along with the annihilation (creation) of a phonon with frequency $\delta=\omega_c -n_c \omega $. 
In this case, $R_e \approx R\, n_p(\delta)$ and $R_a\approx R\, [n_p(\delta)+1]$.  From Eq.~(\ref{eqn:vcpp}) and Fermi's Golden rule for Floquet systems, we can calculate \cite{footnoteSisy}
\begin{align}
R &= 2 \pi  \abs{u_{x n_c}^{z  0}}^2 \mathcal{J}_R({\delta}).
\end{align}
When $\omega > \omega_c/n_c$, a photon is created (annihilated) simultaneously with  a phonon at frequency $-\delta >0$.  In this case, photon emission and absorption are reversed and $R_e=R\,  [n_p(-{\delta})+1]$ and $R_a=R\, n_p(-{\delta})$.  As $\delta$ approaches zero from this side of the resonance and $\eta \to 0$,  the gain rate of the resonator $R_e -R_a = R$ diverges and, at some point, will exceed $\kappa$ and begin lasing.  In this regime, the primary approximation in deriving $g^{(2)}(0)$, that there is no back-action of the resonator field on the DQD, breaks down; however, a full analysis of the saturation mechanisms for this laser (including non-Markovian effects in the phonon bath) is beyond the scope of the present work.  
Despite this instability,  Eq.~(\ref{eqn:rhoc}) still forces the resonator field to satisfy detailed balance until saturation effects take hold, in which case Eq.~(\ref{eqn:rhoc}) is no longer valid.  As a result, we can define an effective temperature on both sides of the resonance 
\be \label{eqn:Teff}
\frac{R_e}{\kappa+R_a}  = e^{-\hbar (\omega_c - n_c \omega)/k_B T_\textrm{eff}}.
\ee
 Our analysis above predicts $T_\textrm{eff}/T = 1$ in these thermal regions.  Figure 2(b) shows $T_\textrm{eff}/T$ in the regions where $g^{(2)}(0)>1.9$ with cavity decay neglected. These calculations include many additional photon creation (annihilation) processes in $R_{e(a)}$ \cite{supp}, but we see that this ratio is still close to one over a large range of $A$ and $\omega$.   In these thermal regions we also observe  an even-odd effect with $n_c$, which arises from the $\sigma_z$ form of the coupling between the resonator and DQD.
When $\epsilon_0=0$, the DQD has to change states every time it exchanges a virtual quantum with the drive, resonator, or phonons. As a result, the thermalizing Raman processes (Fig.~2(b) inset) are suppressed for odd $n_c$ because the total number of virtual processes is odd. For nonzero $\epsilon_0$, this constraint no longer applies and the even-odd effect is weaker.  

The emergence of an {effective} temperature in a sub-system of a non-equilibrium system is a standard phenomena \cite{Casati79}, what is surprising in this case is that this effective temperature is forced to equal the bath temperature.  This indicates that, for small $\delta$, the identification of $\hbar n_c \omega$ with a chemical potential is justified.  The ability to engineer chemical potentials for light, with a temperature controlled by an external bath, has broad utility for quantum simulation with light \cite{Klaers10,Byrnes14,Hafezi14,Hartmann06,Greentree06}

\begin{figure}[t]
\begin{center}
\includegraphics[width = .4 \textwidth]{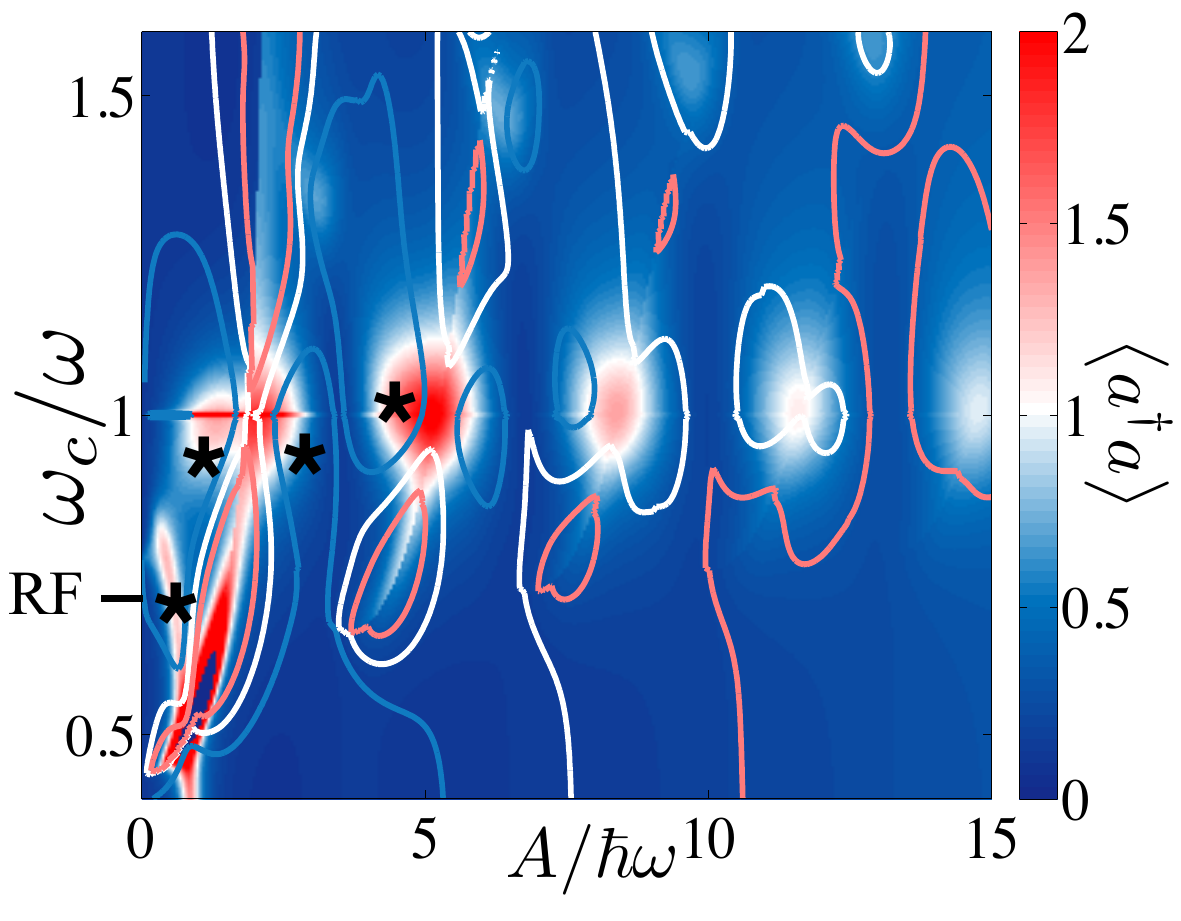}
\caption{Mean photon number $\langle{a^\dagger a}\rangle$ in resonator for varying $A$ and $\omega$ for $\omega_c/2\pi=7.5~$GHz, $\eta=5$~ns$^{-1}$, and other parameters from Fig.~2.  Contours indicate $g^{(2)}(0)$= (1.5/1/0.5) (red/white/blue),  
 asterisks denote single-photon source operating points with $\langle{a^\dagger a}\rangle  \approx 1$ and  $g^{(2)}(0)<0.5$, and RF marks the point of conventional resonance fluorescence $\omega = 2 t_c$.
%
%
}
\label{default}
\end{center}
\end{figure}

%

\emph{Lasing and Single-Photon Source.--}
As shown in Fig.~2(a), away from the thermal resonances, we observe a rich variety of steady state behavior.  
A general feature we observe is oscillations between gain and loss in the resonator transmission with varying drive parameters $A,~\omega,$ and $\epsilon_0$ \cite{supp}.  
In these regions, the gain is not phonon-assisted (as near the thermal resonances) and arises from resonant transitions between Floquet states.  

Figure 2(a) also shows distinct regions where the light in the cavity is strongly antibunched, i.e., $g^{(2)}(0)\ll1$.  This indicates that the DQD-resonator system can act as a microwave single-photon source, similar to what has been achieved with superconducting qubits \cite{Houck07,Wallraff10}.  Ideally one would like to achieve small $g^{(2)}(0)$ and  $\langle{a^\dagger a}\rangle \approx 1$.  As shown in Fig.~3, this is  achieved near the conventional conditions for resonance fluorescence, where $\omega$ is near resonant with the bare two-level system \cite{Kimble76,Kimble77}.  In addition, we find  several other regions at large drive amplitudes where the system also achieves  small $g^{(2)}(0)$ and  $\langle{a^\dagger a}\rangle \approx 1$.  The dynamics in these regions can be understood in terms of resonance fluorescence in the Floquet basis, which has the strongest effect when the the drive frequency and quasi-energy gap are near resonant with the cavity or (not shown) a sub-harmonic of the cavity.

\emph{Conclusions.--}
We  showed that a strongly driven InAs nanowire DQD  can equilibrate the photons in a nearby microwave resonator into a thermal state at the   temperature of the surrounding substrate and a non-zero chemical potential. The highly nonlinear response of the DQD to the drive enables  these chemical potentials to be induced at a harmonic of the drive frequency, allowing for efficient rejection of the drive field.  Outside these thermal regions, we found regimes where the system begins lasing or acts as a microwave single-photon source.  These latter two effects are broadly applicable to other DQD material systems.  Furthermore, one can tune between these diverse regimes \textit{in situ} simply by changing the drive parameters or DQD configuration.
The DQD's broad utility to engineer quantum states of microwave photons suggests many applications to cQED.




\begin{acknowledgements}
\emph{Acknowledgements --} MJG and JMT would like to thank the Kavli Institute for Theoretical Physics where some of this work was completed.  This research was supported in part by the NSF and the NSF Physics Frontier at the JQI.  Research at Princeton was supported by the Gordon and Betty Moore Foundation's EPiQS Initiative through Grant No. GBMF4535, with partial support from the Packard Foundation and the National Science Foundation (Grants No. DMR-1409556 and DMR-1420541).
\end{acknowledgements}

\bibliographystyle{../../../apsrev-nourl}
\bibliography{../../../DDMaser}

\begin{thebibliography}{54}
\bibitem[{foo()}]{footnoteSisy}
\bibinfo{note}{This expression for $R_{e(a)}$ shows that the even-odd effect
  with $n_c$ arises from the matrix element $u_{x n_c}^{z 0}$ and, thus, can be
  considered as a type of selection rule when $\epsilon_0=0$.}
\end{thebibliography}

\pagestyle{empty}
{ 
\begin{figure*}
\vspace{-1.8cm}
\hspace*{-2cm} 
\includegraphics[page=1]{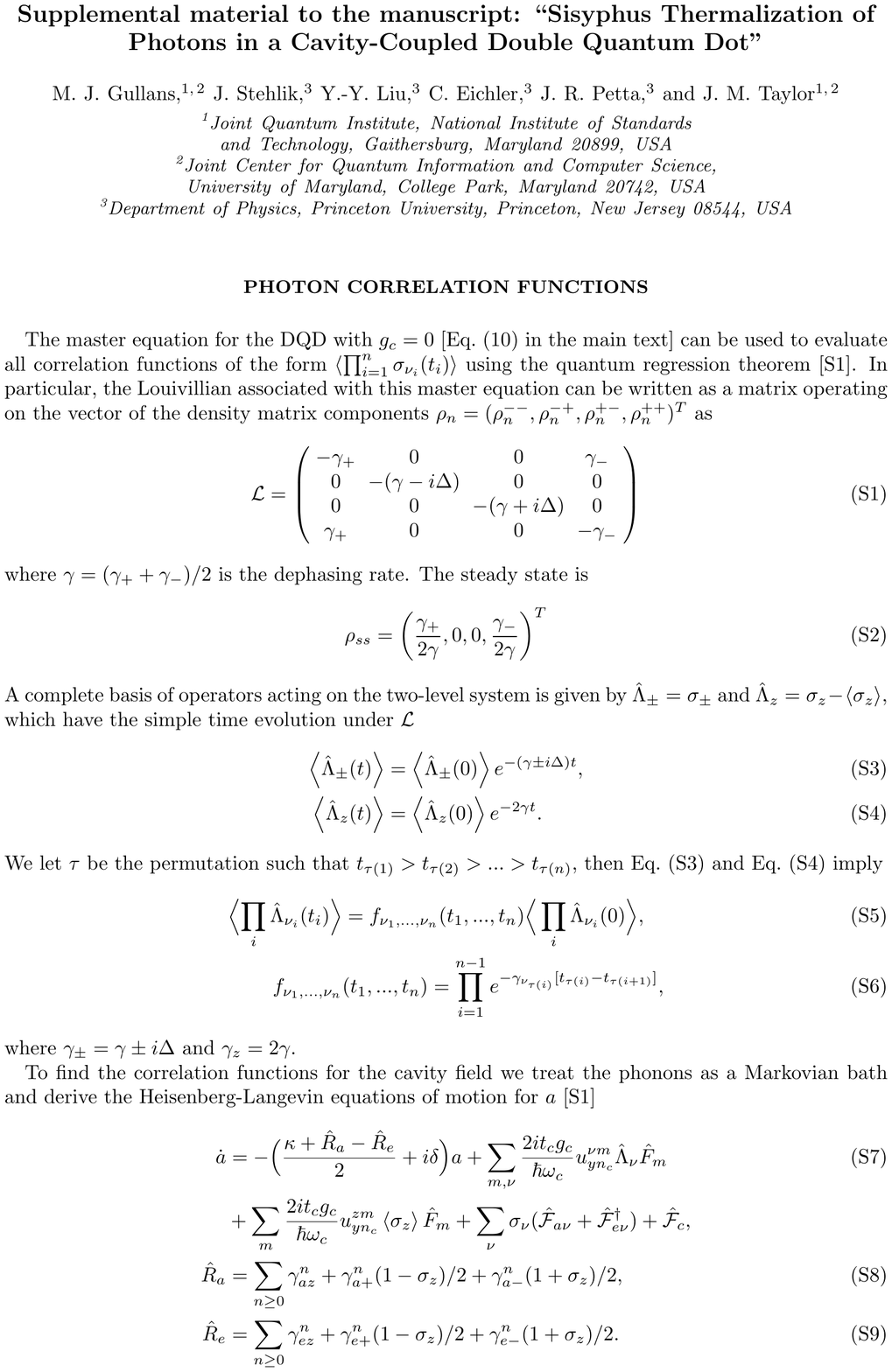}
\end{figure*}

\begin{figure*}
\vspace{-1.8cm}
\hspace*{-2cm} 
\includegraphics[page=2]{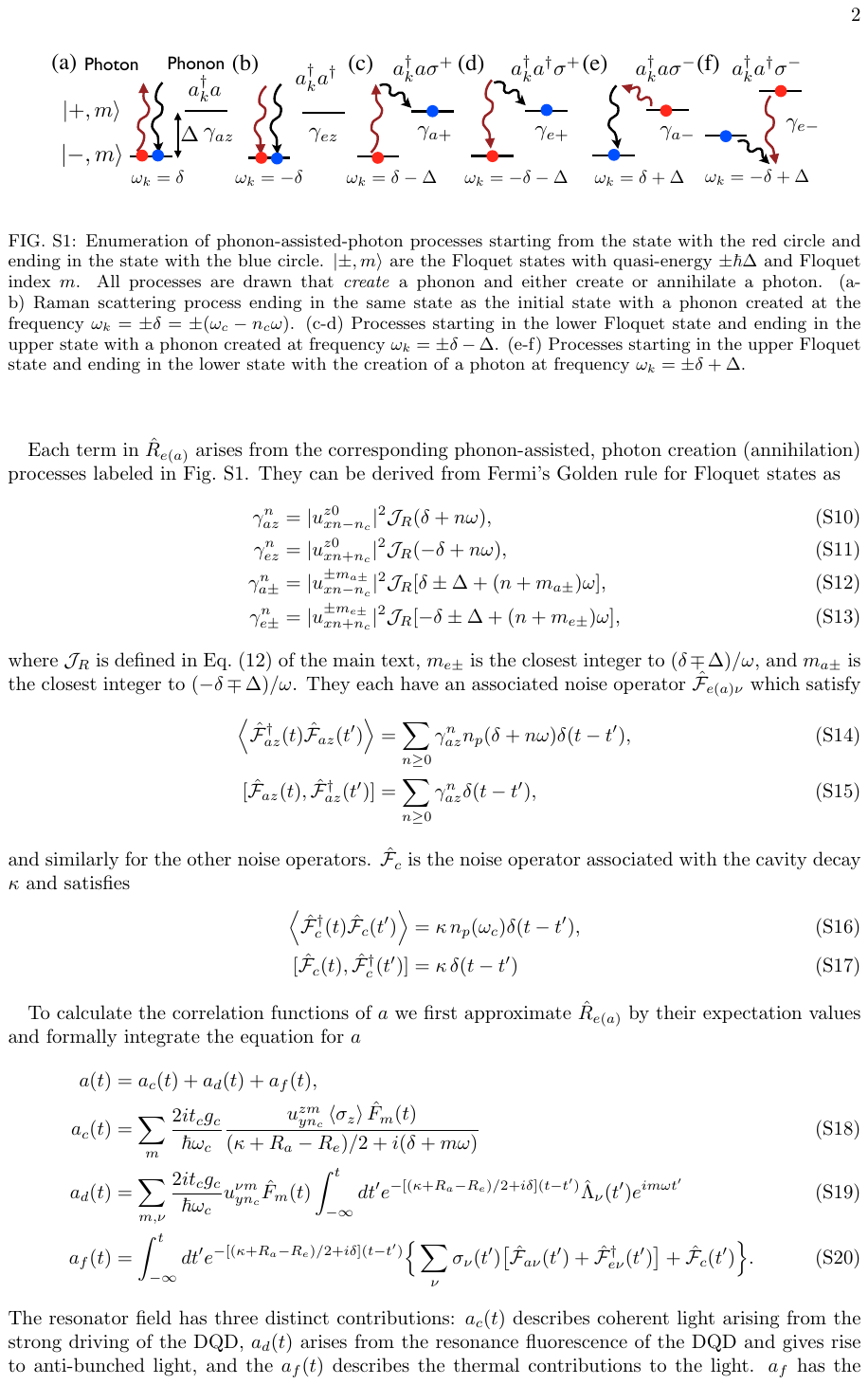}
\end{figure*}

\begin{figure*}
\vspace{-1.8cm}
\hspace*{-2cm} 
\includegraphics[page=3]{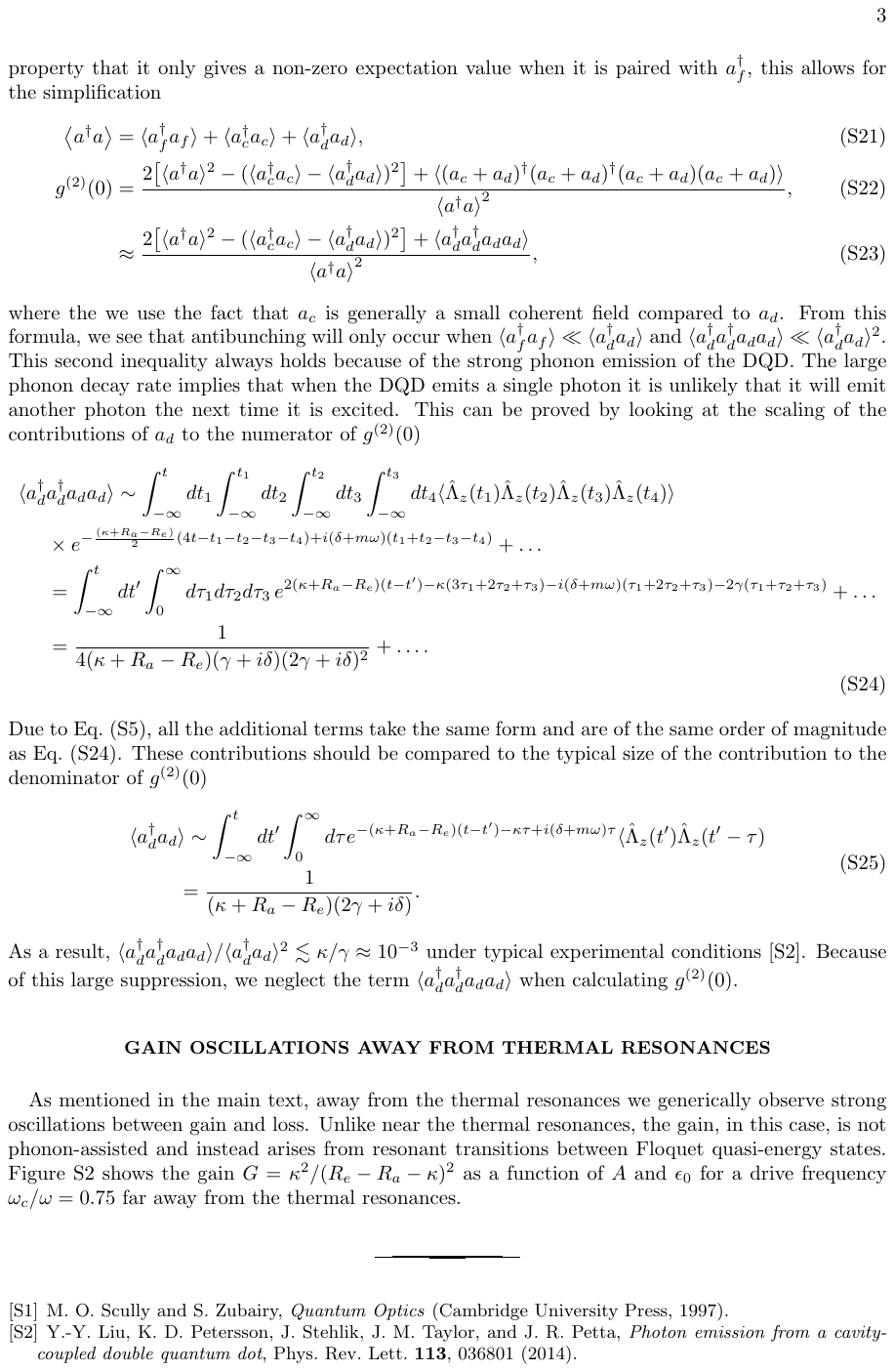}
\end{figure*}

\begin{figure*}
\vspace{-1.8cm}
\hspace*{-2cm} 
\includegraphics[page=4]{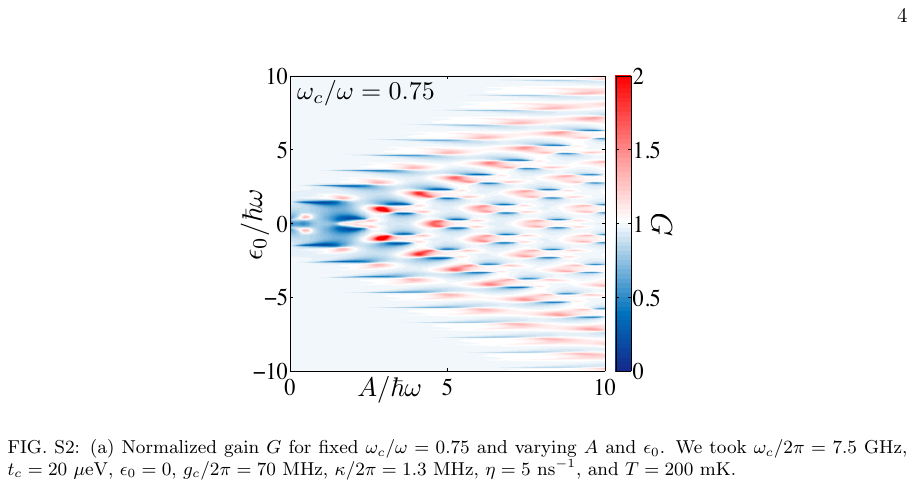}
\end{figure*}

}

\end{document}